# Dual-comb thin-disk oscillator

Kilian Fritsch[1,3,*], Jonathan Brons[1,4], Maksim Iandulskii[1], Ka Fai Mak[2], Zaijun Chen[2], Nathalie Picqué[2] and Oleg Pronin[2,3]

[1] Ludwig-Maximilians-Universität München, Am Coulombwall 1, D-85748 Garching, Germany

[2] Max-Planck-Institut für Quantenoptik, Hans-Kopfermann-Str. 1, D-85748 Garching, Germany

[3] Helmut-Schmidt-Universität / Universität der Bundeswehr Hamburg, Holstenhofweg 85, D-22043 Hamburg, Germany

[4] TRUMPF Laser GmbH, D-78713 Schramberg, Germany

* corresponding author: *kilian.fritsch@hsu-hh.de*

**For the first time to our knowledge, a dual-comb laser based on thin-disk technology and its application to direct frequency comb spectroscopy are presented. The peak power (0.6 MW) and the average power (12 W) of our Yb:YAG thin-disk dual-comb system are more than one-order-of-magnitude higher than in any previous systems. The scheme allows easy adjustment of the repetition frequency difference during operation. A time-domain signal recorded over 10 μs without any active stabilization was sufficient to resolve individual comb lines after Fourier transformation. The demonstration should enable a wider adoption of dual-comb systems towards practical applications in research laboratories. Its simplicity and compactness especially for the realization of tri- and multi-comb systems and conversion into the still poorly covered UV and VUV ranges makes it a promising next-generation technology.**

A Fourier transform spectrometer [1] measures the time-domain interference between two waves while their relative phase-shift is varied. The Fourier transform of the interference pattern provides the spectrum. Advanced implementations use lasers as light sources. Their increased brightness compared to incoherent sources leads to shorter measurement time or increased signal-to-noise ratio [2]. Even further refined systems incorporate optical frequency combs (OFC) as light sources. This enhances the spectral resolution, ultimately limited by the OFCs individual comb linewidth, improves sensitivity [3] and reduces the effect of instrument line shape on the measurement [2]. Experiments comprising two OFCs, known as dual-comb spectroscopy (DCS), emulate the mechanically scanned delay in Fourier-transform spectroscopy by two separate pulse trains with slightly different repetition frequencies (asynchronous pulse trains). DCS is rapidly advancing since its first proposal [4] and demonstration in a Ti:Sapphire-driven mid-infrared system [5]. The key reason is the simple and fast electronic detection of the interferogram generated by two heterodyned asynchronous pulse trains on a photodetector or by means of electro optic sampling [6]. Advantages arise in terms of acquisition speed, spectral resolution, accuracy, robustness and signal-to-noise ratio (SNR) over traditional scanning Fourier-transform spectroscopy. The principle is extendable to nearly all spectroscopy methods and at present several fundamental experimental demonstrations were published including Doppler-free spectroscopy [7], anti-Stokes Raman- [8] and two-photon spectroscopy [9] as well as scattering scanning near-field optical microscopy [10,11]. The applications include trace gas detection, lidars, kinetics of chemical reactions, as reviewed in [2][12].

One major challenge with DCS is the implementation of an interferometer with two combs. They need to be mutually coherent and operate at different repetition rates [12], whilst spanning the necessary bandwidth in the examined spectral region. One promising way of realizing such a light source is by electronically modulating a continuous wave laser with two intensity modulators at different frequencies [13]. More commonly though, the two OFCs are generated by separate mode-locked femtosecond oscillators running at slightly offset repetition rates, which involves the complex active stabilization of repetition rates and carrier-envelope offset frequencies of both or their active tracking and a-posteriori correction. To date, erbium fiber lasers [14] show the best performance based on active stabilization or a-posteriori correction schemes.

An emerging way is to generate a dual-comb output from a single laser cavity. This could potentially simplify the technique, as many noise sources are expected to be shared by both pulse trains. This alleviates the need for active stabilization over short time periods. So far, there have been only a few demonstrations of such a scheme. They rely on different effects to generate a repetition frequency offset between both outputs. For example, polarization multiplexing in MIXSELs [15] and chromatic dispersion in a dual-wavelength fiber laser system [16] was demonstrated. In bulk solid-state lasers, the directional dependence of the Kerr nonlinearity [17] was exploited. A selection of the passive dual-comb lasers is given in supplementary table 1. It compares average and peak power, repetition rate, difference in repetition rates, pulse duration, spectral span, central wavelength and type of laser technology. Average and peak powers are particularly important for spectral broadening, pulse compression and spectral extension into the THz, infrared (IR), ultraviolet (UV) and vacuum ultraviolet (VUV) spectral regions where most of the strong and spectroscopically relevant transitions are situated. The conversion efficiency, which is intrinsically low for these nonlinear processes, needs to be compensated by a high-photon flux driving laser.

In this work we present, to the best of our knowledge, the first thin-disk based dual-comb femtosecond oscillator. Thanks to the thin-disk platform [18], the presented system shows over an order of magnitude higher output average and peak powers compared to the previous dual-comb demonstrations. No noise is added by external optically active media, since the thin-disk laser provides enough power without subsequent amplification. The scheme effectively rejects common noise without any active stabilization. We report a broad tuning range of the difference of the repetition frequencies $\Delta f_{\text{rep}}$ during mode-locked laser operation by simply adjusting the relative longitudinal offset of the end mirrors. In contrast to the aforementioned optical systems based on a single cavity in supplementary table 1, this allows optimal and flexible choice of the important parameter $\Delta f_{\text{rep}}$.

Given its output parameters, it represents a relatively simple and compact table-top source for DCS, particularly promising for further nonlinear conversion into the IR, UV and VUV spectral regions and realization of different nonlinear spectroscopy schemes.

**Experimental Setup**

The experimental setup (Fig. 1) is a high power Kerr-lens mode-locked (KLM) Yb:YAG thin-disk oscillator pumped by a 500 W-class fiber-coupled diode laser. It features a thin-disk module consisting of the disk and the pump cavity made by TRUMPF as well as high power optimized dispersive mirrors designed and manufactured by Dr. V. Pervak and his team. The dual layout of this thin-disk laser contains two nearly identical laser resonators which share all cavity mirrors including the active area on the thin-disk (Fig. 1c, pump spot). An exception to this principle constitutes the end mirrors and output couplers. They are split horizontally to allow the alignment of both cavities independently [19]. The spatial separation is roughly 5 mm in the vertical (sagittal) plane (Fig. 1a, separation drawn horizontally (tangentially) for simplified visualization). The spatial separation allows individual fine tuning of the repetition frequencies. Their difference $\Delta f_{\text{rep}}$ is a key parameter in any DCS systems. It determines the dead-time between consecutive interferograms and the line-spacing of the down-sampled RF comb [20]. In the presented arrangement, this can be easily tuned up to the kilohertz range by longitudinally translating one of the split output couplers. This simple tunability of $\Delta f_{\text{rep}}$ can allow the implementation of fast scanning methods such as those relying on the inversion of the sign of $\Delta f_{\text{rep}}$, demonstrated in [21]. This is a major advantage over other DCS systems relying on different methods to introduce $\Delta f_{\text{rep}}$, i.e. via birefringence [22] or chromatic dispersion [16]. Usually those are quite limited in the tunability and tunability range (see supplementary table 1 for more information), because changes will involve rearranging or realigning of the cavities as opposed to simply translating a mirror.

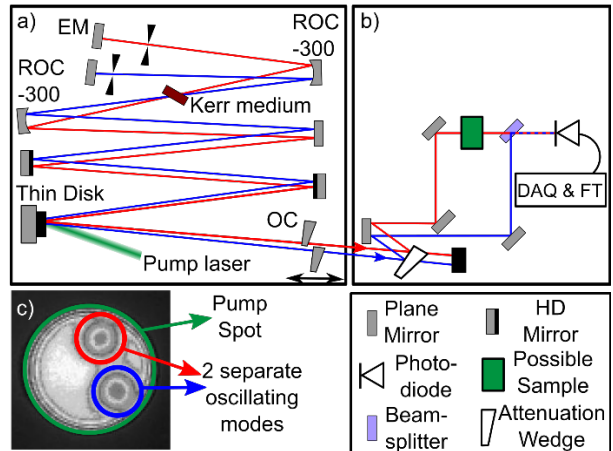

**Fig. 1.** Schematic drawing of the dual-comb thin-disk laser system. a) The laser is based on the Yb:YAG thin-disk resonator. The beam-paths corresponding to the "red" and "blue" beams are separated in tangential plane for illustration only. EM – end mirror; ROC – radius of curvature; OC – output coupler (8% transmission) b) The heterodyne detection setup. DAQ – data acquisition board; FT – Fourier transform. c) Image of the pump spot (laser active area) on the thin disk. Both laser modes share a single pump spot and therefore a single pump laser

The configuration of an individual cavity, without the stacked dual-cavity feature, was inspired by Brons, et. al. [23]. A telescope comprising two concave mirrors with a radius of curvature (ROC) of 300 mm focuses the cavity modes into a 3 mm-thin sapphire plate under Brewster's angle, which acts as a Kerr-nonlinear medium. Radial restriction of the modes by hard apertures and soft aperture effects in the gain medium combined with the self-focusing in the sapphire plate enables KLM [24]. Despite the vertical displacement of the laser beams, it is possible to use a single sapphire plate to mode-lock both cavities. Similar to conventional KLM thin-disk oscillators, each cavity can be mode-locked by manually perturbing one of the telescope mirrors. Surprisingly, this process also works for both lasers simultaneously. In order to achieve soliton mode-locking in the net-negative dispersion regime, two highly dispersive mirrors, compensating a Group Delay Dispersion of approximately -3000 fs$^2$ per reflection, resulting in - 12 000 fs$^2$ total intra-cavity dispersion per round-trip, are placed inside the resonators. The common-path cavity design ensures temporal stability of the dual –output system as well as a reasonable mutual coherence time between the two frequency combs in absence of any active stabilization mechanisms. Both cavities share nearly the same intra-cavity optical components and the same pump source. Thus, they experience nearly equal perturbations originating from air flows, mechanical vibrations and the intensity noise of the pump source.

**Output characteristics of the dual-comb system**

The spectra of both pulse trains, measured with an optical spectrum analyzer (OSA, Ando), are compared in Fig. 2. They show almost identical spectral shapes and widths, indicating almost identical output pulses. Both spectra are centered at 1030 nm and have slightly different average output powers of 14 W and 12 W, which corresponds to 230 nJ and 200 nJ per pulse, respectively. This corresponds to a combined optical-to-optical efficiency of 11% at the 240 W pump power. The repetition frequency $f_{rep}$ was set to 61 MHz. The sech²-shaped output pulses have 300 fs and 305 fs FWHM pulse durations, measured with a commercial autocorrelator (APE). The peak powers are calculated to be around 0.58 MW and 0.67 MW, respectively. To the best of our knowledge, the achieved peak power exceeds the currently reported highest peak power from a single laser dual-comb system of 0.03 MW [17] by approximately 20 times. It is also more than 9 times larger than the currently reported highest peak power of 0.07 MW from a dual-comb CPA system [25]. The relatively small differences between the two outputs can arise from the fact that the optical axes are not identical resulting in the different angles of incidence on the telescope mirrors and Kerr medium. It can be mitigated easily by attenuating the more powerful beam to match the lower power one. The different beam positions on the thin-disk can also result in a slight gain difference between the two lasing modes. Unequal amounts of defects on the mirrors, as well as a mismatch in the size of the hard apertures can also lead to additional differences in the linear losses.

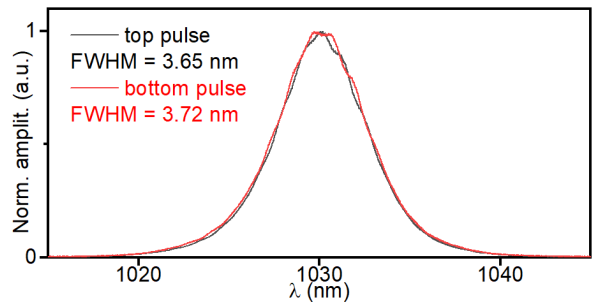

**Fig. 2.** Spectral characteristics of the emitted pulses.

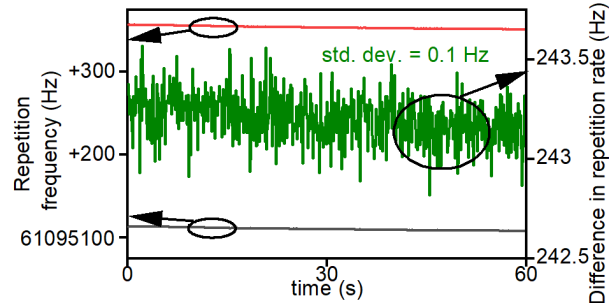

**Fig. 3.** Temporal stability of the repetition rates and the difference in the repetition rates. Both repetition frequencies monotonically decrease with time (black and red curves).

The time evolution of the two repetition frequencies and their difference is shown in Fig. 3. Although both repetition frequencies decrease monotonically with time, their difference $\Delta f_{\text{rep}}$ fluctuates only weakly with a standard deviation of 0.1 Hz on a minute time-scale. This may hint at $\Delta f_{\text{rep}}$ being constant on a milli- to microsecond time scale, the typical acquisition times for fast DCS experiments. The presented stability level is comparable to previous publications 16,17,26,27, which is remarkable considering the at least one order of magnitude higher output average powers and intra-cavity power of approximately 150 W. We therefore conclude that this stability level should suffice for fast DCS experiments and potentially be acceptable for slower measurements over several seconds. Further active stabilization of the difference in repetition frequencies should not be required, but could be implemented for more challenging experiments. Furthermore, the pulse energy is over an order of magnitude higher compared to 7.7 nJ used from the CPA system in [25] and far exceeds the energies delivered by state-of-the-art single laser systems. In future versions of the presented system, the pulse energy could be scaled up easily as described by Brons, et. al. [18] through increasing focal length of the cavities curved mirrors. For example, a set of -600 mm mirrors instead of -300 mm mirrors should yield a twofold increase.

**Dual-Comb Measurements**

By combining two laser beams on a photodiode (see setup in Fig. 1b), interferograms are recorded in the time domain. Fig. 4a shows a 50 ms trace with 9 consecutive interferograms. $\Delta f_{\text{rep}}$ was adjusted to 189 Hz, which corresponds to 5.3 ms temporal separation between consecutive interferograms and a down-conversion factor of $a = \frac{\Delta f_{\text{rep}}}{f_{\text{rep}}} = 3.09 * 10^{-6}$ [20]. The optical spectrum can be obtained by taking a Fourier transform of an individual interferogram centered in a 20 µs time-window (Fig. 4c) and scaling it appropriately by the down-conversion factor. No sample was used in this measurement. The spectrum agrees well with a reference spectrum recorded by an OSA (Ando).

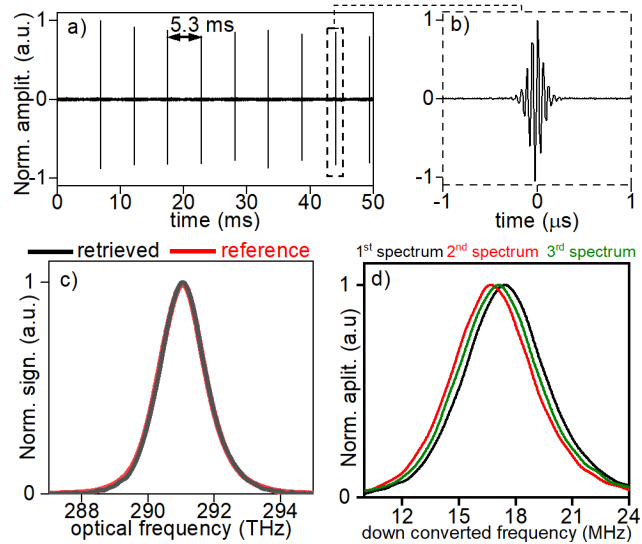

**Fig. 4.** Measurements with the dual-comb system without any sample. a) Time-trace with 9 consecutive interferograms. Time separation between neighboring interferograms is 5.3 ms. b) Single interferogram. c) Single spectrum retrieved from a single 20 µs time-trace and the reference spectrum. d) Three spectra obtained from three consecutive interferograms. All spectra have a frequency shift relatively to each other.

In order to investigate how the interferograms differ from one another, each of them was analyzed and Fourier transformed individually within the 20 µs time-window around its peak. Fig. 4d depicts three spectra, retrieved from three consecutive interferograms. The spectra show identical shapes while being offset in the frequency domain. As mentioned before, $\Delta f_{\text{rep}}$ remains constant on the millisecond timescale and therefore cannot affect the width of the spectra during the measurement. This may indicate that the dominating instabilities are due to the relative carrier-envelope offset frequency ($\Delta f_{\text{CEO}}$) fluctuations, which can be compensated numerically by means of a simple post-processing technique (see Supplement 1 for detail). In order to resolve the individual comb-lines, the measurement time needs to be increased to yield enough data to enhance frequency resolution. Furthermore, the increase of the acquisition time and averaging will improve the signal-to-noise ratio (SNR) if the two frequency combs stay coherent.

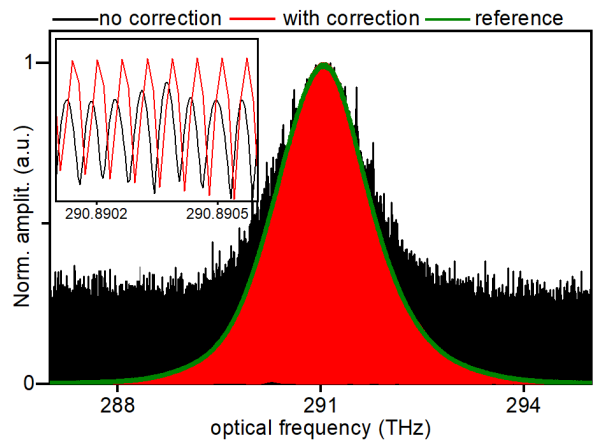

**Fig. 5.** Phase-corrected (red curve) and not corrected (black curve) spectra of 10 ms time-trace. Two interferograms contributed in both cases. It is clearly seen that phase-corrected spectrum experiences significant noise decrease. The inset shows the frequency comb lines of both spectra.

Fig. 5 depicts optical spectra calculated from a trace containing two interferograms. A comparison between the spectra with the already mentioned phase correction technique applied and the spectra without any correction shows that the SNR for the phase-corrected case is significantly improved while the good agreement with the reference measurement remains unchanged. The inset of Fig. 5 shows that the spectral resolution is already sufficiently high to resolve the comb structure of the optical spectrum. The 8 modes with the mode-spacing of 61 MHz span a band of 3.6 GHz.

Furthermore, a time-trace consisting of 465 consecutive interferograms was analyzed applying the phase-correction. The analysis showed that the FWHM of a single comb line (Fig. 6) of the spectrum is 0.35 Hz in the RF domain. Interleaving multiple measurements can bring the spectral resolution close to this value [28]. Moreover, using the aforementioned phase-correction technique on arbitrarily long measurements containing a vast number of interferograms has the potential to improve the spectral resolution even further. The mutual coherence time between the two frequency combs is estimated to be around 1 ms, approximately equal to the meta-stable life time of $Yb^{3+}$ ion in the YAG host material [29].

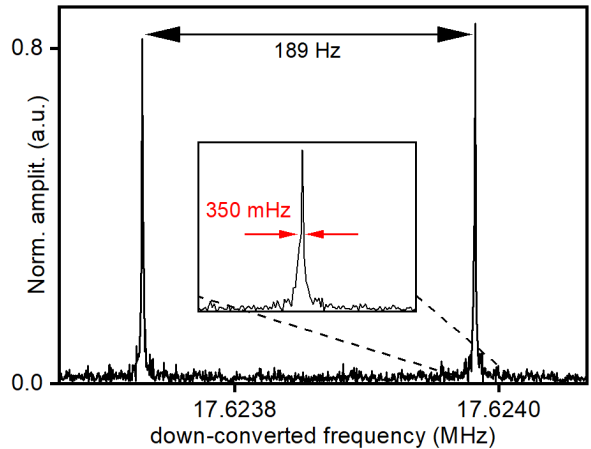

Fig. 6. Comb line structure of the optical spectrum. The inset shows a single comb line with FWHM of 0.35 Hz in radio frequency domain.

Additionally, we performed a spectroscopic measurement of the transmission of a home-built Fabry-Pérot etalon with a free spectral range of 120 GHz and a FWHM-linewidth of 12 GHz. The reference spectrum (OSA) and the measured (retrieved) spectrum are shown in Fig. 7. The retrieved spectrum was acquired from a single 50 µs interferogram time-trace. The characteristic Fabry-Pérot transmission lines are clearly visible and both spectra agree very well. Thus, the single 50 µs interferogram time-trace leads to the spectral resolution of at least 12 GHz.

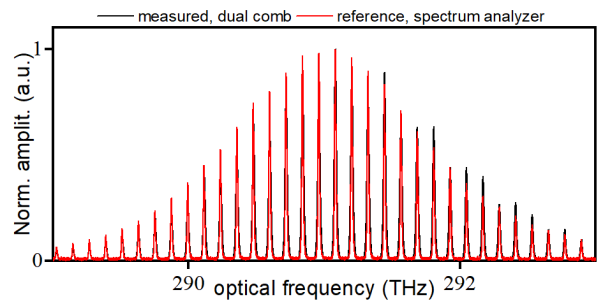

Fig. 7. Comparison of the Fabry-Pérot etalon spectrum measured with the OSA and dual-comb setup. The dual-comb spectrum was obtained from a single 50 µs interferogram time-trace. The width of the transmission line is 12 GHz.

## Discussion

The passive dual-comb oscillator relies on the thin-disk technology with its intrinsic power scalability. Hence, this first demonstration should be able to approach parameters of the state-of-the-art single output systems of around 62 MW peak- and 270 W average powers [18,23]. The presented dual-comb cavity layout is independent of the mode-locking mechanism. Therefore it's adoption to powerful SESAM mode-locked systems [30][31] is not out of scope. Moreover, the disk geometry with its large, flat and easily scalable active area allows for a simple integration of multiple spatial modes. Thus, not only the dual output systems but multiple output lasers can be demonstrated. Multiple output oscillators would open a way towards very compact, simple, cost effective multi-comb, for example tri-comb, systems for 2D spectroscopy [32] and generally nonlinear multi-dimensional spectroscopy and imaging.

The significant limitation, which strongly influences the data retrieval, is the loss of coherence between interferograms due to $\Delta f_{CEO}$ fluctuations. It was shown that a simple phase-correction technique can compensate those fluctuations sufficiently well for the presented measurements. One very attractive approach for the passive $f_{CEO}$ stabilization was already demonstrated with the KLM thin-disk oscillators by means of intra-pulse difference frequency generation and simultaneous broadband conversion in mid-IR range at ~100 mW average power [33]. The required spectral broadening and temporal pulse compression were also demonstrated for a large span of peak and average powers in fibers or multipass cells [34]. Furthermore, a system driven by a mode-locked thin-disk Ho:YAG oscillator running at 2.1 μm wavelength [35] showed excellent conversion efficiencies [36] and ultra-broadband coverage (500 to 2250 cm$^{-1}$) of the IR region. Therefore, a combination of the passive dual-comb approach and the Ho:YAG thin-disk technology should be possible in the near future. This would open a route towards ultra-broadband dual-comb spectroscopy of diverse chemical and biological samples [37]. Interestingly, the systems can also be upgraded with an electro-optic sampling unit [6] avoiding the use of liquid nitrogen detectors.

However, the extension of the spectral range towards UV and VUV might be even more interesting as no dual-comb lasers are available in this spectral range [2]. Efficient spectral broadening or nonlinear frequency conversion to UV (257 nm) is already possible at the current power level and will be even simpler for the power scaled systems. The mutual coherence time estimated to be around 1 ms should be sufficient to preserve good passive coherence while recording one single interferogram after nonlinear conversion in VUV. Yet, this topic clearly requires further investigation, since the required relative stability scales with frequency and the harmonic generation will add a significant amount of noise.

It is worth noting that novel methods developed for CEO stabilization of KLM thin-disk oscillators can also be applied to the dual-comb systems to actively stabilize $\Delta f_{CEO}$ [38,39]. This might be necessary once the amount of interferograms becomes too large to be numerically a posteriori phase-corrected. Direct averaging of the interferograms in a FPGA (Field-programmable gate array) can be implemented instead. The stabilization of $\Delta f_{CEO}$ is expected to be simpler as compared to $f_{CEO}$ solely due to its easier detection.

## Conclusion

In summary, we have developed a passive dual frequency comb spectrometer based on a single free-running Yb:YAG KLM thin-disk laser. To the best of our knowledge, the peak (0.6 MW) and average power (12 W) is an order of magnitude higher compared to all other dual-comb systems. Both cavities emit nearly identical pulse trains, mutually coherent on a time scale of approximately 1 ms. This experimentally-achieved coherence is well manifested during one single interferogram showing comb lines in the spectrum after the Fourier transformation without any active stabilization and phase corrections. The passive stability of the difference in repetition rates was measured to be around 0.1 Hz over 60 s. This value is sufficient for most of the experiments involving dual frequency combs. It was shown that, in combination with the simple phase-correction technique, a narrow comb line width of 350 mHz in the RF domain (corresponding to 113 kHz in the optical domain) has been achieved, which sets the ultimate resolution of this spectrometer for interleaving measurements [28].

Ultrafast thin-disk oscillators have been demonstrated with average powers approaching 300 W and peak powers as high as 62 MW. The power scalability of our platform and plethora of different developments of this technology mentioned above, opens up new strategies for all the applications of frequency comb spectroscopy. Extension towards UV and VUV ranges seems especially promising.

## Acknowledgements

We thank Prof. F. Krausz for his insightful remarks and strong support of this research as well as Dr. V. Pervak and his team for designing and manufacturing the optimized dispersive mirror coating.

## Disclosures

The authors declare no conflicts of interest.

Dual-comb thin-disk oscillator - K. Fritsch, et. al.    12

# Reference

1. P. R. Griffiths and J. A. de Haseth, *Fourier transform infrared spectrometry, second edition*, 2nd ed. (John Wiley & Sons, 2007).

2. N. Picqué and T. W. Hänsch, "Frequency comb spectroscopy," Nature Photon **13**, 146–157 (2019).

3. J. Mandon, G. Guelachvili, and N. Picqué, "Fourier transform spectroscopy with a laser frequency comb," Nature Photon **3**, 99–102 (2009).

4. S. Schiller, "Spectrometry with frequency combs," Opt. Lett. **27**, 766 (2002).

5. F. Keilmann, C. Gohle, and R. Holzwarth, "Time-domain mid-infrared frequency-comb spectrometer," Opt. Lett. **29**, 1542 (2004).

6. A. Bartels, R. Cerna, C. Kistner, A. Thoma, F. Hudert, C. Janke, and T. Dekorsy, "Ultrafast time-domain spectroscopy based on high-speed asynchronous optical sampling," The Review of scientific instruments **78**, 35107 (2007).

7. S. A. Meek, A. Hipke, G. Guelachvili, T. W. Hänsch, and N. Picqué, "Doppler-free Fourier transform spectroscopy," Optics letters **43**, 162–165 (2018).

8. T. Ideguchi, S. Holzner, B. Bernhardt, G. Guelachvili, N. Picqué, and T. W. Hänsch, "Coherent Raman spectro-imaging with laser frequency combs," Nature **502**, 355 EP - (2013).

9. A. Hipke, *Dual-Frequency-Comb Two-Photon Spectroscopy*, Dissertation, Ludwig-Maximilians-Universität, 2016.

10. S. Amarie, T. Ganz, and F. Keilmann, "Mid-infrared near-field spectroscopy," Optics express **17**, 21794–21801 (2009).

11. M. Brehm, A. Schliesser, and F. Keilmann, "Spectroscopic near-field microscopy using frequency combs in the mid-infrared," Optics express **14**, 11222–11233 (2006).

12. I. Coddington, N. Newbury, and W. Swann, "Dual-comb spectroscopy," Optica **3**, 414 (2016).

13. G. Millot, S. Pitois, M. Yan, T. Hovhannisyan, A. Bendahmane, T. W. Hänsch, and N. Picqué, "Frequency-agile dual-comb spectroscopy," Nature Photon **10**, 27 EP - (2016).

14. Z. Chen, M. Yan, T. W. Hänsch, and N. Picqué, "A phase-stable dual-comb interferometer," Nature communications **9**, 3035 (2018).

15. S. M. Link, D. J. H. C. Maas, D. Waldburger, and U. Keller, "Dual-comb spectroscopy of water vapor with a free-running semiconductor disk laser," Science (New York, N.Y.) **356**, 1164–1168 (2017).

16. X. Zhao, G. Hu, B. Zhao, C. Li, Y. Pan, Y. Liu, T. Yasui, and Z. Zheng, "Picometer-resolution dual-comb spectroscopy with a free-running fiber laser," Optics express **24**, 21833–21845 (2016).

17. T. Ideguchi, T. Nakamura, Y. Kobayashi, and K. Goda, "Kerr-lens mode-locked bidirectional dual-comb ring laser for broadband dual-comb spectroscopy," Optica **3**, 748 (2016).

18. J. Brons, V. Pervak, E. Fedulova, D. Bauer, D. Sutter, V. Kalashnikov, A. Apolonskiy, O. Pronin, and F. Krausz, "Energy scaling of Kerr-lens mode-locked thin-disk oscillators," Optics letters **39**, 6442–6445 (2014).

# Supplementary Material: Dual-comb thin-disk oscillator

Kilian Fritsch[1,3], Jonathan Brons[1,4], Maksim Iandulskii[1], Ka Fai Mak[2], Zaijun Chen[2], Nathalie Picqué[2] and Oleg Pronin[2,3]

[1] Ludwig-Maximilians-Universität München, Am Coulombwall 1, D-85748 Garching, Germany

[2] Max-Planck-Institut für Quantenoptik, Hans-Kopfermann-Str. 1, D-85748 Garching, Germany

[3] Helmut-Schmidt-Universität / Universität der Bundeswehr Hamburg, Holstenhofweg 85, D-22043 Hamburg, Germany

[4] TRUMPF Laser GmbH, D-78713 Schramberg, Germany, Amtsgericht Stuttgart HRB 747703

*Corresponding author: *kilian.fritsch@hsu-hh.de*

**Phase-correction for longer time-traces**

To numerically compensate changes in $\Delta f_{\mathrm{CEO}}$, a simple phase-correction technique, knows as Forman-Vanasse phase-correction [1], was implemented. All interferograms were centered within a few µs time-windows (e.g. 2 µs in case of the phase-corrected spectrum in Fig. 5) and extracted from the original time trace (FigS. 1a). Then, all these windowed interferograms were Fourier transformed. As was shown in Fig. 4d, the spectra from consecutive interferograms are frequency shifted related to each other. All these spectra can be realigned by means of frequency correction (FigS. 1b): all spectra are centered to the reference value, which can be, for example, the peak frequency of the first spectrum. Since all spectra are now shifted in the frequency domain, according to the shift property of the Fourier transform [2], all corresponding windowed interferograms should be multiplied by a phase factor $e^{2\pi i t f_{\mathrm{shift}}}$:

$$h(t) = e^{2\pi i t f_{\mathrm{shift}}} g(t), \text{ then: } H(t) = G(f - f_{\mathrm{shift}})$$

where $H, G$ — the Fourier transforms of $h(t)$ and $g(t)$ respectively. The phase factor depends linearly on the time and the frequency shift $f_{\mathrm{shift}}$. The above mentioned phase-correction allows to reduce the noise in the retrieved spectrum and illuminate the difference between the interferograms due to the $\Delta f_{\mathrm{CEO}}$ change. The signal can be further reconstructed by stitching all the windowed corrected interferograms together (FigS. 1c) inserting a zero-signal between them to fill in the original time separation between them. Since all information is carried in the interferogram signal, which lasts for only several microseconds in this specific experimental setup, inserting a zero-signal between the phase-corrected interferograms can reduce noise impact to spectrum. This specific step of the post-processing method is only applicable to measurements without a sample. Otherwise time between burst, ideally, will be filled with signal emitted by the sample. Some parts of the time-trace outside interferogram can be occupied by background noise. The spectrum (Fig.S 2a) which is taken from 10 ms time-trace indicates that the signal itself is rather noisy. All these high-peak noise components arise from surrounding noise sources inside a lab and were numerically filtered (FigS. 2b, Fig. 5).

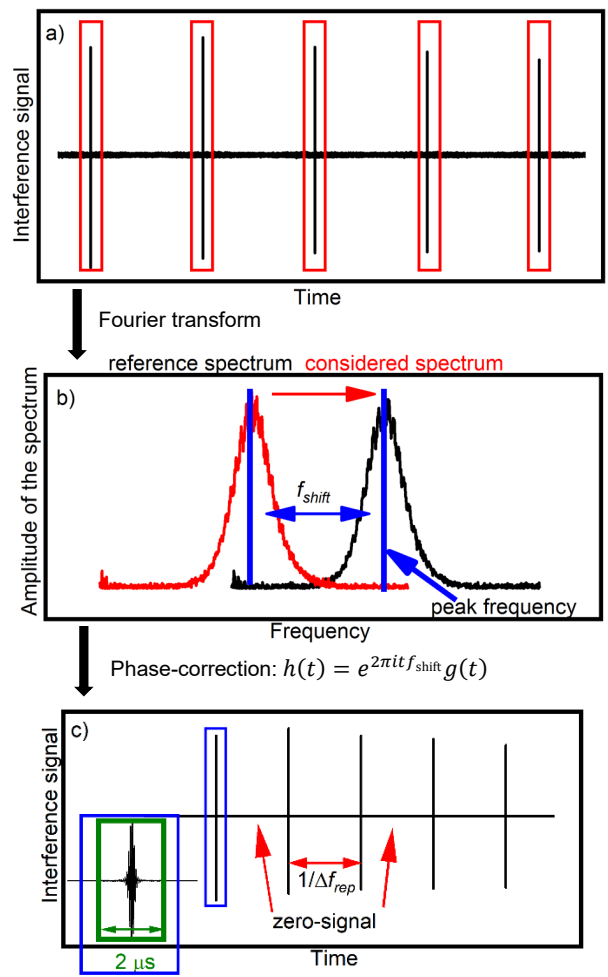

**FigS. 1.** Schematic representation of the post processing phase-correction procedure (not to scale). a) - Windowing of the interferograms; b) - Frequency correction of the spectra and further phase-correction of the windowed interferograms; c) - Stitching of all

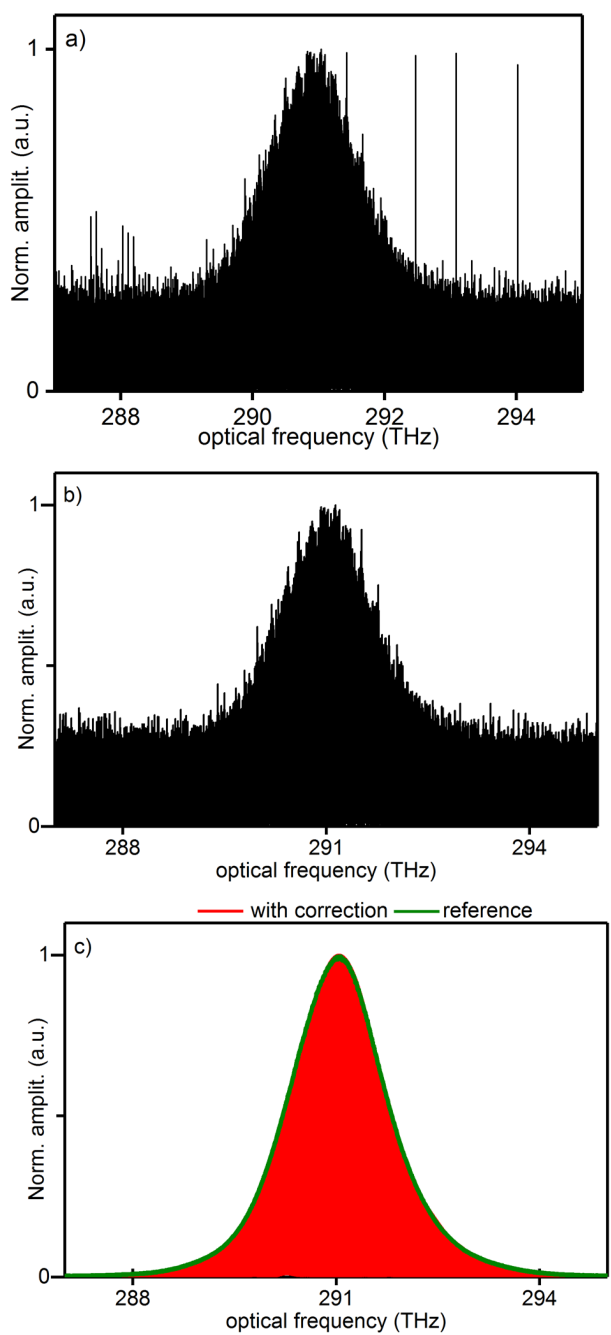

**FigS. 2.** a) Spectrum obtained from 10 ms time-trace. Two interferograms contributed. High-peak noise components originate from external noise sources inside a lab. b) Spectrum obtained from 10 ms time-trace without high-peak noise components. c) Phase-corrected spectrum obtained from 10 ms time-trace. Two

# Overview State-of-the-art

**Supplementary table 1: Non-complete selection as overview of dual-comb laser systems with high power (HP) and / or single-cavity (SC) properties.**
(___) Underlined $\Delta f_{rep}$ values can be easily adjusted by the user during operation of the system.
(~) values are not directly given in the publication but were estimated from given parameters

| PUBLICATION | | $f_{REP}$ [MHZ] | $\Delta f_{REP}$ | $E_{PULSE}$ | $P_{AVG}$ | $P_{PEAK}$ | $\tau_{PULSE}$ | $\lambda_0$ [NM] | $\Delta\lambda$ [NM] | $\Delta\nu$ [THZ] |
|---|---|---|---|---|---|---|---|---|---|---|
| THIS WORK | SC | 61.1 | Up to kHz | 190 nJ | 12 W | 0.7 MW | 300 fs | 1030 | 3.7 | 1.1 |
| | HP | | | 230 nJ | 14 W | 0.6 MW | | | | |
| AKOSMAN ET AL. [3] | SC | 67.6 | 510 Hz | ~89 pJ | 6 mW | ~0.2 kW | 400 fs | 1975 | 9.8 | ~0.75 |
| ZHAO ET AL. [4] | SC | 52.7 | 1.3 kHz | ~0.3 nJ | 16 mW | ~2.7 kW | ~105 fs | 1533 | 33 | ~4.2 |
| | | | | ~0.5 nJ | 25 mW | ~2.8 kW | ~160 fs | 1544 | 22 | ~2.8 |
| LINK ET AL. [5] | SC | 1700 | 4 MHz | ~35 pJ | 60 mW | ~2 W | 18 ps | 968.3 | ~0.1 | ~0.03 |
| IDEGUCHI ET AL. [6] | SC | 932 | 325 Hz | ~360 pJ | 340 mW | ~29 kW | 12 fs | 837 | 61 | 26.1 |
| | | | | ~330 pJ | 310 mW | ~26 kW | | | | |
| NÜRNBERG ET AL. [7] | HP | 2730 | 51.9 kHz | ~8.8 pJ | 24 mW | ~21 W | 400 fs | 1030 | 2.8 | ~0.8 |
| | | | | ~10.3 pJ | 28 mW | ~24 W | | | 3.6 | ~1.0 |
| MILLOT ET AL. [8] | HP | 300 | 100 kHz | 16 pJ | 50 mW | 3.33 W | ~5 ps | 1530 - 1625 | 2 | 0.37 |
| | | | | 83 pJ | 250 mW | 16.75 W | ~5 ps | | 3 | |
| LIAO ET AL. [9] | SC | 71.9 | 3.2 kHz | ~41 pJ | 3 mW | 0.14 kW | 270 fs | 1917 | 35 | ~2.9 |
| | | | | | | 0.16 kW | 250 fs | 1981 | 20 | ~1.5 |
| NAKAJIMA ET AL. [10] | SC | 37.9 | 1.5 Hz | ~48 pJ | 1.8 mW | 0.6 kW | ~75 | 1550 | 56 | 7.0 |
| | | | | ~32 pJ | 1.2 mW | 0.4 kW | | | | |
| IDEGUCHI ET AL. [11] | HP | 100 | 100 Hz | 13 nJ | 1.3 W | ~600 kW | 20 | 795 | ~47 | 22 |
| BERNHARDT ET AL. [12] | HP | 130 | 600 Hz | ~7.7 nJ | 1 W[1] | ~72.4 kW | 100 | 1040 | 30 | 8.3 |
| MEHRAVAR ET AL. [13] | SC | 72.4 | 82 Hz | ~0.07 nJ | 5.1 mW | ~175 W | ~370 | 1555 | 9.6 | 1.2 |
| OLSON ET AL. [14] | SC | 60.8 | 229 Hz | ~15 pJ | 0.9 mW | ~56 W | ~260 | 1865 | 20 | 1.72 |
| MOHLER ET AL. [15] | HP | 1000 | 2 kHz | ~0.5 nJ | 0.5 W | ~23 kW | 20 | 795 | 60 | 28 |

---

[1] The System is capable of 17 W average power, but was used at only 1 W